\documentclass[twocolumn,showkeys,aps,prb,showpacs]{revtex4-1}
\UseRawInputEncoding
\usepackage{graphicx}
\usepackage[CJKbookmarks,dvipdfm,colorlinks,linkcolor=red,citecolor=blue]{hyperref}

\begin{document}

\title{Importance of magnetic shape anisotropy in determining magnetic and electronic properties of monolayer $\mathrm{VSi_2P_4}$}

\author{San-Dong Guo$^{1}$,  Yu-Ling Tao$^{1}$,  Kai Cheng$^{1}$, Bing Wang$^{2}$ and Yee-Sin Ang$^{3}$}
\affiliation{$^1$School of Electronic Engineering, Xi'an University of Posts and Telecommunications, Xi'an 710121, China}
\affiliation{$^2$Institute for Computational Materials Science, School of Physics and Electronics,Henan University, 475004, Kaifeng, China}
\affiliation{$^3$Science, Mathematics and Technology (SMT), Singapore University of Technology and Design (SUTD), 8 Somapah Road, Singapore 487372, Singapore}
\begin{abstract}
Two-dimensional (2D) ferromagnets  have been a fascinating subject of research, and magnetic anisotropy (MA)
is indispensable for stabilizing the 2D magnetic order. Here, we investigate  magnetic anisotropy energy (MAE), magnetic and electronic properties of   $\mathrm{VSi_2P_4}$ by using the generalized gradient approximation plus $U$ (GGA+$U$) approach. For large $U$,  the magnetic shape anisotropy (MSA) energy  has a more pronounced contribution to the MAE, which can overcome the magnetocrystalline anisotropy (MCA) energy to evince an easy-plane.
For fixed out-of-plane MA,  monolayer $\mathrm{VSi_2P_4}$ undergoes  ferrovalley (FV), half-valley-metal (HVM), valley-polarized quantum anomalous Hall insulator (VQAHI), HVM and FV states with increasing $U$.  However, for  assumptive in-plane MA, there is no special quantum anomalous Hall (QAH) state and spontaneous valley polarization within considered $U$ range. According to the MAE and electronic structure with fixed out-of-plane or in-plane MA, the intrinsic phase diagram shows common magnetic semiconductor (CMS), FV and VQAHI in monolayer $\mathrm{VSi_2P_4}$. At representative $U$$=$3 eV widely used in references, $\mathrm{VSi_2P_4}$ can be regarded as a 2D-$XY$ magnet, not Ising-like 2D long-range order magnets predicted in previous works with only considering MCA energy.
Our findings shed light on importance of MSA in determining magnetic and electronic properties of monolayer $\mathrm{VSi_2P_4}$.

\end{abstract}
\keywords{Magnetic anisotropy, 2D ferromagnets, Phase transition  ~~~~~~~~~~~~~Email:sandongyuwang@163.com}

\maketitle

\section{Introduction}
There has been a tremendous interest in searching  2D magnetic materials, which can interplay with other important properties of materials such as FV, ferroelectricity, piezoelectricity and QAH effects.
According to the Mermin-Wagner theorem,  the
long-range magnetic order at finite temperature for the 2D isotropic Heisenberg spin systems is prohibited\cite{a1}.
Experimentally, both
the $\mathrm{CrI_3}$ monolayer  and $\mathrm{Cr_2Ge_2Te_6}$ bilayer  show an
easy out-of-plane magnetization\cite{a1-1,a1-2}, while the 2D magnet $\mathrm{CrCl_3}$
has an easy in-plane magnetization\cite{a1-3,a1-4}. The underlying mechanism is that MA
is responsible for  the stable 2D magnetic order.
With the presence of out-of-plane magnetization, the long-range FM order can sustain
at finite temperature by opening a magnon gap to resist thermal
agitations\cite{a1-5}.  An easy-plane anisotropy can  realize the quasi-ordered  topological vortex/antivortex pairs, which can be described by the
Berezinskii-Kosterlitz-Thouless (BKT) theory based on the 2D $XY$ model\cite{a1-6}.
Moreover, MA can affect the symmetry of 2D systems, and then produce important influence on their topological and valley properties\cite{a4,a5,a6,a7}, when the spin-orbital coupling (SOC) is included.
For these 2D systems, the FV and QAH states can exist with fixed out-of-plane magnetic anisotropy, but these special states disappear for in-plane case\cite{a5,a6,a7}.

The magnetization direction can be determined by MAE, which mainly include MCA and MSA energies.  Compared with SOC-induced MCA,
 the MSA due to the magnetic dipole-dipole (D-D) interaction is often relatively weak and is neglected.
But for a 2D system with  weak MCA, the MSA may have an important contribution. The representative example is 2D magnet $\mathrm{CrCl_3}$, which has been confirmed to have an easy in-plane magnetization in experiment\cite{a1-3,a1-4}. When only considering MCA energy, the $\mathrm{CrCl_3}$  shows an
easy out-of-plane magnetization, which is contrary to experimental results\cite{a1-7}. When including MSA energy, the theoretical results agree well with  experimental in-plane magnetization\cite{a1-7}.  Another case is  CrSBr monolayer\cite{a7-1}.   The MCA energy  shows
that the $b$ axis is the easy one, and the $a$ axis would be hard.  When including MSA energy,  the MAE gives the easy
magnetization $b$ axis and hard $c$ axis.

\begin{figure*}
  \includegraphics[width=15cm]{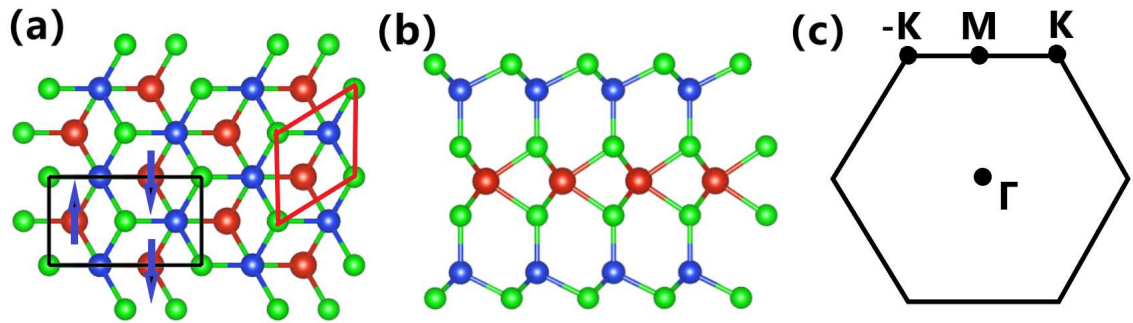}
  \caption{(Color online)For $\mathrm{VSi_2P_4}$ monolayer,  (a): top view and (b): side view of  crystal structure. The primitive (rectangle supercell) cell is
   shown by red (black) lines, and the AFM configuration is marked with  blue arrows in (a). (c): the first BZ with high symmetry points. }\label{st}
\end{figure*}

\begin{figure}
  \includegraphics[width=8cm]{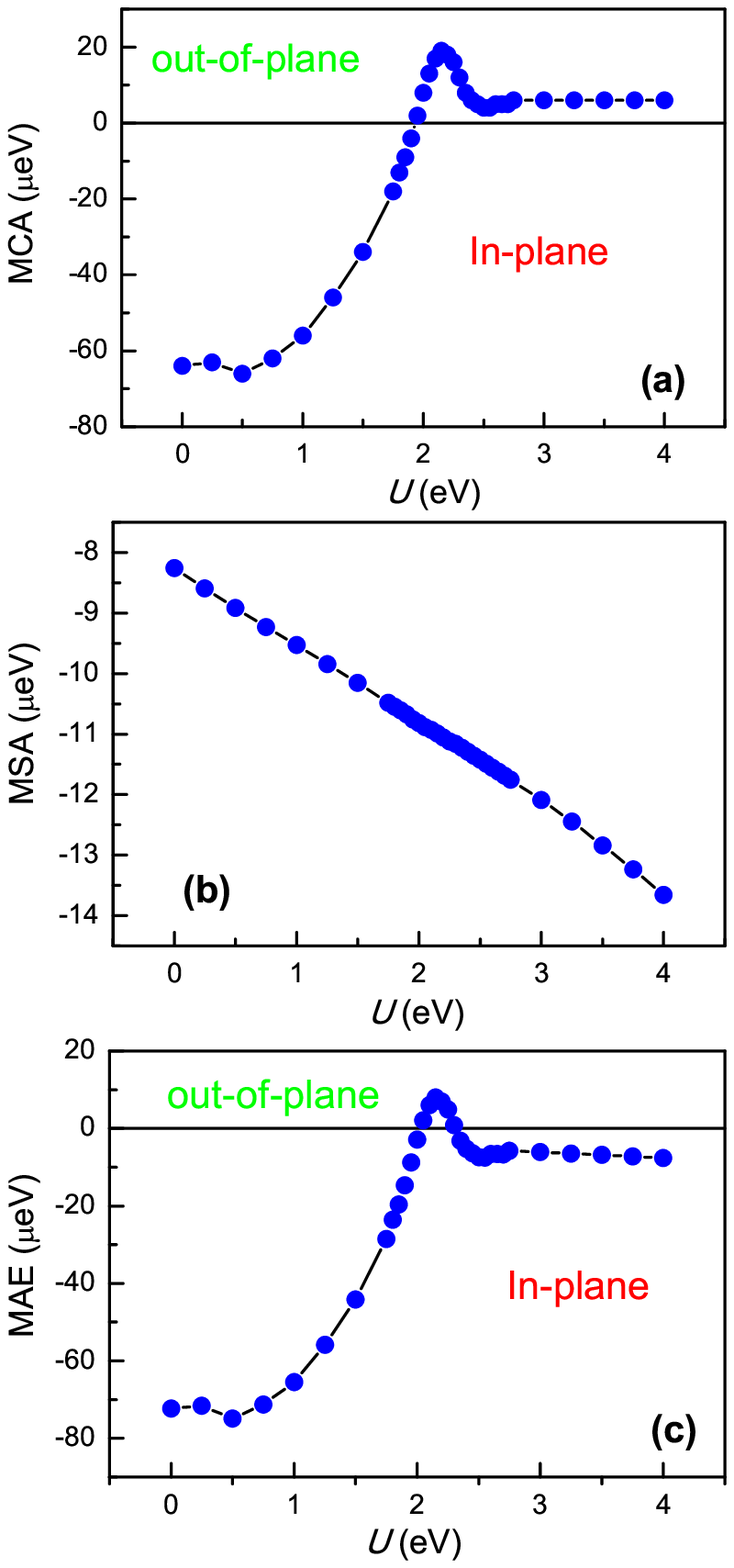}
  \caption{(Color online) For $\mathrm{VSi_2P_4}$ monolayer, MCA energy (a), MSA energy (b) and MAE (c) as a function of $U$. }\label{mae}
\end{figure}

Recently, the septuple-atomic-layer
2D $\mathrm{MoSi_2N_4}$ and $\mathrm{WSi_2N_4}$ have been  successfully
synthesized  by the chemical vapor deposition
method\cite{a11}. Subsequently,  2D $\mathrm{MA_2Z_4}$
family  with a septuple-atomic-layer structure  has been established, which possess emerging topological,
magnetic, valley and superconducting properties\cite{a12}. The $\mathrm{VA_2Z_4}$ of them possess magnetic properties, which have been widely investigated\cite{a5,a1-8,a1-9,a1-10,a1-10,a1-11,a1-12,a1-13} due to their FV properties. The monolayer $\mathrm{VSi_2P_4}$ is  a typical example, which is predicted to have out-of-plane magnetization with only considering MCA energy (6.38 $\mathrm{\mu eV}$)\cite{a1-9}. In recent works, the Janus monolayer $\mathrm{VSiGeN_4}$ is predicted to be in-plane magnetization with only considering MCA energy, and the in-plane magnetization is enhanced, when including MSA energy (about -17 $\mathrm{\mu eV}$)\cite{a1-10,a1-13}. This implies that $\mathrm{VSi_2P_4}$ may be an in-plane 2D $XY$ magnet without  the long-range FM order.
In this work, we investigate MCA energy, MSA energy, MAE and electronic structures of $\mathrm{VSi_2P_4}$ as a function of $U$,  and reveal the
importance of MSA  in determining its magnetic, topological and valley properties.
At representative $U$$=$3 eV\cite{a5,a12,a1-8,a1-9,a1-10,a1-10,a1-11,a1-12,a1-13}, $\mathrm{VSi_2P_4}$ can be regarded as a 2D-$XY$ magnet,  which  is different from previous conclusion that the Ising-like 2D long-range order magnet is predicted  with only considering MCA energy\cite{a5,a1-9}.

The rest of the paper is organized as follows. In the next
section, we shall give our computational details and methods.
 In  the next few sections,  we shall present structure and MAE, and electronic structures   of  $\mathrm{VSi_2P_4}$ monolayer. Finally, we shall give our conclusion.

\section{Computational detail}
Based on density-functional
theory  (DFT)\cite{1}, the spin-polarized  first-principles calculations  are performed  by employing the projected
augmented wave method,  as implemented in VASP code\cite{pv1,pv2,pv3}.
We use popular generalized gradient approximation of Perdew-Burke-Ernzerhof (PBE-GGA)\cite{pbe}  as exchange-correlation functional.
To consider on-site Coulomb correlation of V  atoms,  the GGA+$U$  method is used within  the rotationally invariant approach proposed by Dudarev et al\cite{u}.
To attain accurate results, we use the energy cut-off of 500 eV,  total energy  convergence criterion of  $10^{-8}$ eV and  force
convergence criteria of less than 0.0001 $\mathrm{eV.{\AA}^{-1}}$ on each atom.
A vacuum space of more than 32 $\mathrm{{\AA}}$ is used to avoid the interactions
between the neighboring slabs.
We use $\Gamma$-centered 15 $\times$15$\times$1 k-point meshs  in  the Brillouin zone (BZ) for structure optimization and electronic structures calculations, and 9$\times$16$\times$1 Monkhorst-Pack k-point meshs for calculating ferromagnetic (FM)/antiferromagnetic (AFM)  energy  with rectangle supercell.
 The SOC effect is explicitly included to investigate MCA, electronic and topological properties of  $\mathrm{VSi_2P_4}$ monolayer.
The Berry curvatures  are calculated  directly from  wave functions  based on Fukui's
method\cite{bm}, as implemented in VASPBERRY code\cite{bm1,bm2}.
The mostly localized Wannier functions  are constructed from the $d$-orbitals of V atom, $s$-and $p$-orbitals of Si and Ge,  $p$-orbitals of N atoms by Wannier90 code\cite{w1}, and then edge states are calculated by employing
the  WannierTools package\cite{w2}.

\section{Structure and magnetic anisotropy energy}
The crystal structures of monolayer   $\mathrm{VSi_2P_4}$ are plotted  in \autoref{st}, along with  the first BZ with high-symmetry points.
The primitive cell contains one V, two Si, and four P atoms, which
is stacked by seven atomic layers of P-Si-P-V-P-Si-P. Each V atom
is coordinated with six P atoms, which forms  a trigonal prismatic
configuration.  In other words,  this $\mathrm{VP_2}$ layer is sandwiched by two Si-P bilayers, and the crystal symmetry of
 $\mathrm{VSi_2P_4}$ is $P\bar{6}m2$ (No.187) with broken inversion symmetry.
The optimized  lattice constants $a$ of $\mathrm{VSi_2P_4}$ monolayer is 3.486 $\mathrm{{\AA}}$,  agreeing well with previous
theoretical value\cite{a5}.

Based on Goodenough-Kanamori-Anderson rules\cite{q18,q18-1},
the superexchange coupling between two V atoms  is FM, because the V-P-V  bonding angle is $91.68^{\circ}$, which  is  close to $90^{\circ}$.
To confirm this, the energy differences (per formula unit) between AFM and FM ordering as a function of $U$ are also plotted in FIG.1 of electronic supplementary information (ESI).
Calculated results show that the FM state is always the magnetic ground state of $\mathrm{VSi_2P_4}$ within  considered $U$ range.
It is found that energy difference between  AFM and FM ordering around  $U$=1.45 eV has a sudden jump, which is due to abrupt change of  magnetic moment of  V atom for  AFM ordering (The magnetic moment of  V atom changes from 0.74 $\mu_B$ to 0 $\mu_B$  to 0.079 $\mu_B$ to 0.81 $\mu_B$ , when $U$ varies from 1.3 eV to 1.4 eV to 1.5 eV to 1.6 eV),  and the small magnetic moment of  V atom  reduces  the magnetic interaction energy.   Similar phenomenon can be observed, when energy difference between  AFM and FM ordering  is as a function of strain for $\mathrm{VSiGeN_4}$ and $\mathrm{VSi_2P_4}$\cite{a1-9,a1-13}.

The  orientation of magnetization plays an important role on magnetic and electronic states of some 2D materials\cite{a4,a5,a6,a7}.
The MAE can be used to measure the dependence of the energy on the orientation of magnetization.
The MAE  originates mainly from two parts: (1) MCA energy $E_{MCA}$, which is an intrinsic property of the material caused by
the SOC; (2) MSA energy $E_{MSA}$), which is basically
the anisotropic D-D interaction\cite{a1-7,a7-1}:
 \begin{equation}\label{d-d}
E_{D-D}=\frac{1}{2}\frac{\mu_0}{4\pi}\sum_{i\neq j}\frac{1}{r_{ij}^3}[\vec{M_i}\cdot\vec{M_j}-\frac{3}{r_{ij}^2}(\vec{M_i}\cdot\vec{r_{ij}})(\vec{M_j}\cdot\vec{r_{ij}})]
 \end{equation}
where the $\vec{M_i}$ represents the local magnetic moments and $\vec{r_{ij}}$
are vectors that connect the sites $i$ and $j$.
 For the monolayer with a collinear FM,  when the spins lie in
the plane, the \autoref{d-d} can be written as:
 \begin{equation}\label{d-d-1}
E_{D-D}^{||}=\frac{1}{2}\frac{\mu_0M^2}{4\pi}\sum_{i\neq j}\frac{1}{r_{ij}^3}[1-3\cos^2\theta_{ij}]
 \end{equation}
 where $\theta_{ij}$ is the angle between the $\vec{M}$ and $\vec{r_{ij}}$.   When the spins are out of plane ($\theta_{ij}$=$90^{\circ}$), the \autoref{d-d} can further be simplified as:
 \begin{equation}\label{d-d-2}
E_{D-D}^{\perp}=\frac{1}{2}\frac{\mu_0M^2}{4\pi}\sum_{i\neq j}\frac{1}{r_{ij}^3}
 \end{equation}
The $E_{MSA}$  ($E_{D-D}^{||}-E_{D-D}^{\perp}$)can be  expressed as:
\begin{equation}\label{d-d-3}
E_{MSA}=\frac{3}{2}\frac{\mu_0M^2}{4\pi}\sum_{i\neq j}\frac{1}{r_{ij}^3}\cos^2\theta_{ij}
 \end{equation}
The MSA tends to make magnetic moments of magnetic elements directed parallel to the surfaces, which can minimize
magnetostatic energy.  Usually, the MCA energy dominates the MAE, and MSA energy can be ignored. However, MSA becomes significant to MAE for
materials with weak SOC.

Firstly, we calculate MCA energy  by $E_{MCA}=E^{||}_{SOC}-E^{\perp}_{SOC}$, which is obtained from
GGA+$U$+SOC calculations. The $E_{MCA}$ as a function of $U$ is plotted in \autoref{mae} (a). Calculated results show  multiple transitions in the MCA.
With increasing $U$, the  $E_{MCA}$ changes from negative value to positive value, which means a transition from in-plane to out-of-plane.
At $U$=3.0 eV, the calculated   $E_{MCA}$ (about 6 $\mathrm{\mu eV}$) is consistent with available result (6.38 $\mathrm{\mu eV}$)\cite{a1-9}.
Secondly, we calculate MCA energy  by \autoref{d-d-3}, which depends on the crystal structure and local magnetic moment of V ($M_V$).
The $M_V$ as a function of $U$ is shown in FIG.2 of ESI. It is found that $M_V$ increases (0.89 $\mu_B$-1.15 $\mu_B$) with increasing $U$.
 The $E_{MSA}$ as a function of $U$ is shown in \autoref{mae} (b), and the $E_{MSA}$ changes from -8 $\mathrm{\mu eV}$ to -14 $\mathrm{\mu eV}$, when $U$ varies from 0 eV to 4 eV. Calculated results show that the MSA is in favour of in-plane anisotropy. Finally, the MAE can be obtained by  $E_{MAE}$=$E_{MCA}$+$E_{MSA}$, which is plotted in  \autoref{mae} (c). When $U$ is between 2.03 eV and 2.31 eV, the positive MAE means a preferred out-of-plane polarization. When $U$$>$2.31 eV and $U$$<$2.03 eV, the in-plane anisotropy can be observed due to negative MAE.

 For $\mathrm{VA_2Z_4}$, the $U$=3 eV has been adopted in previous calculations\cite{a5,a12,a1-8,a1-9,a1-10,a1-10,a1-11,a1-12,a1-13}. The $\mathrm{VSi_2P_4}$ is predicted to be an out-of-plane ferromagnet (Ising-like 2D magnets) due to only considering MCA energy\cite{a1-9}. However, our calculated results show that  $\mathrm{VSi_2P_4}$ can be regarded as a 2D-$XY$ magnet like the $\mathrm{CrCl_3}$ monolayer\cite{a1-3,a1-4,a1-7}. The in-plane easy magnetization means that
there is no energetic barrier to the rotation of magnetization in the $xy$ plane, and a BKT magnetic transition to a quasi-long-range
phase can be observed at a critical temperature\cite{re5,re5-1}. The critical temperature $T_{BKT}=1.335\frac{J}{K_B}$ is obtained from the results of
the Monte Carlo simulation for 2D-XY magnets on a triangular lattice\cite{re6,re7}, where $J$ is the nearest-neighboring exchange parameter and $K_B$ is the Boltzmann
constant.  In a FM configuration, each V has six
neighbors with the same spin, while, for AFM configuration, V has four
neighbors with the opposite spin, and  has two with the same spin. Based on the Heisenberg spin model, the  FM  ($E_{FM}$) and AFM ($E_{AFM}$) energy can be written as:
\begin{equation}\label{pe0-1-2}
E_{FM}=E_0-(6J+2A)S^2
 \end{equation}
  \begin{equation}\label{pe0-1-3}
E_{AFM}=E_0+(2J-2A)S^2
 \end{equation}
where $E_0$ is the total energy of systems without magnetic coupling, and $A$ means the easy-axis single-ion anisotropy.
Hence, $J$ can be obtained directly from the energy difference:
 \begin{equation}\label{pe0-1-3}
J=\frac{E_{AFM}-E_{FM}}{8S^2}
 \end{equation}
The calculated $J$ is 16.07 meV ($S=\frac{1}{2}$) at $U$=3 eV, and the $T_{BKT}$  is predicted to be  249 K.

\begin{figure}
  \includegraphics[width=7.5cm]{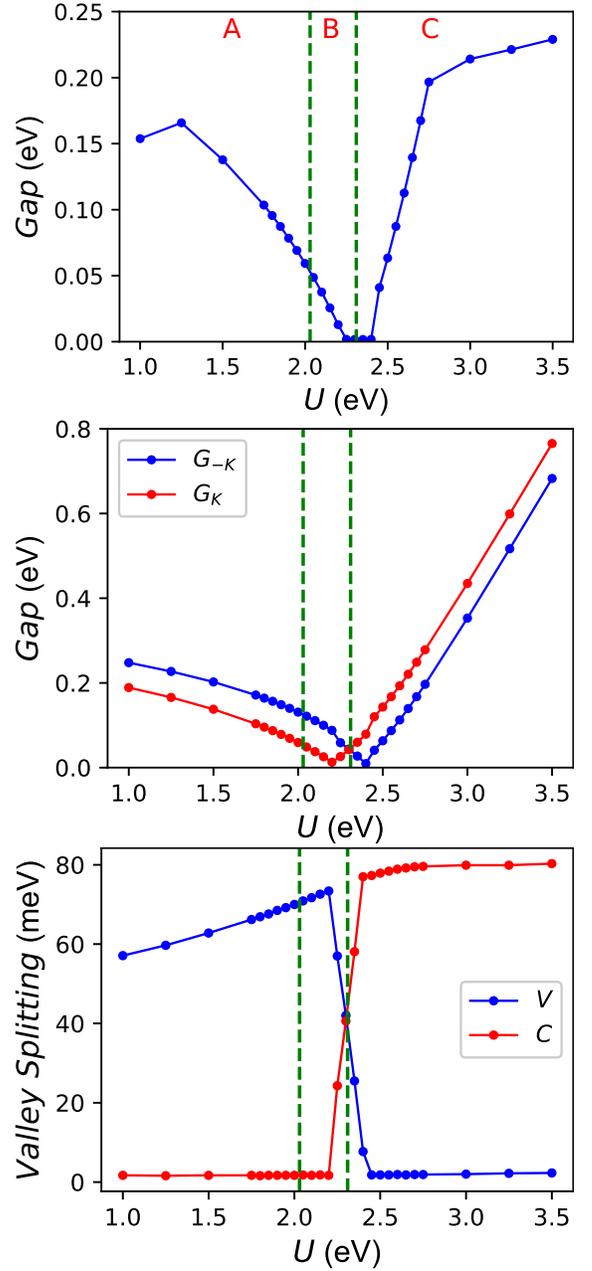}
  \caption{(Color online)For $\mathrm{VSi_2P_4}$ monolayer with fixed out-of-plane  MA, Top panel:  the  global energy band gap; Middle panel: the energy  band gaps for -K and K valleys; Bottom panel: the valley splitting for both valence and condition bands  as a function of   $U$. The A and C regions mean in-plane, while the B region for out-of-plane. }\label{gap}
\end{figure}
\begin{figure}
  \includegraphics[width=8cm]{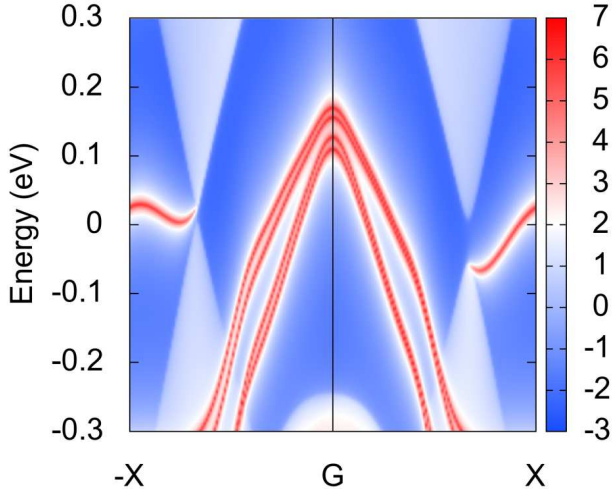}
  \caption{(Color online)For $\mathrm{VSi_2P_4}$ monolayer with out-of-plane  MA,  the topological
edge states  at representative  $U$$=$2.25 eV. }\label{s-s}
\end{figure}

\begin{figure}
  \includegraphics[width=7.5cm]{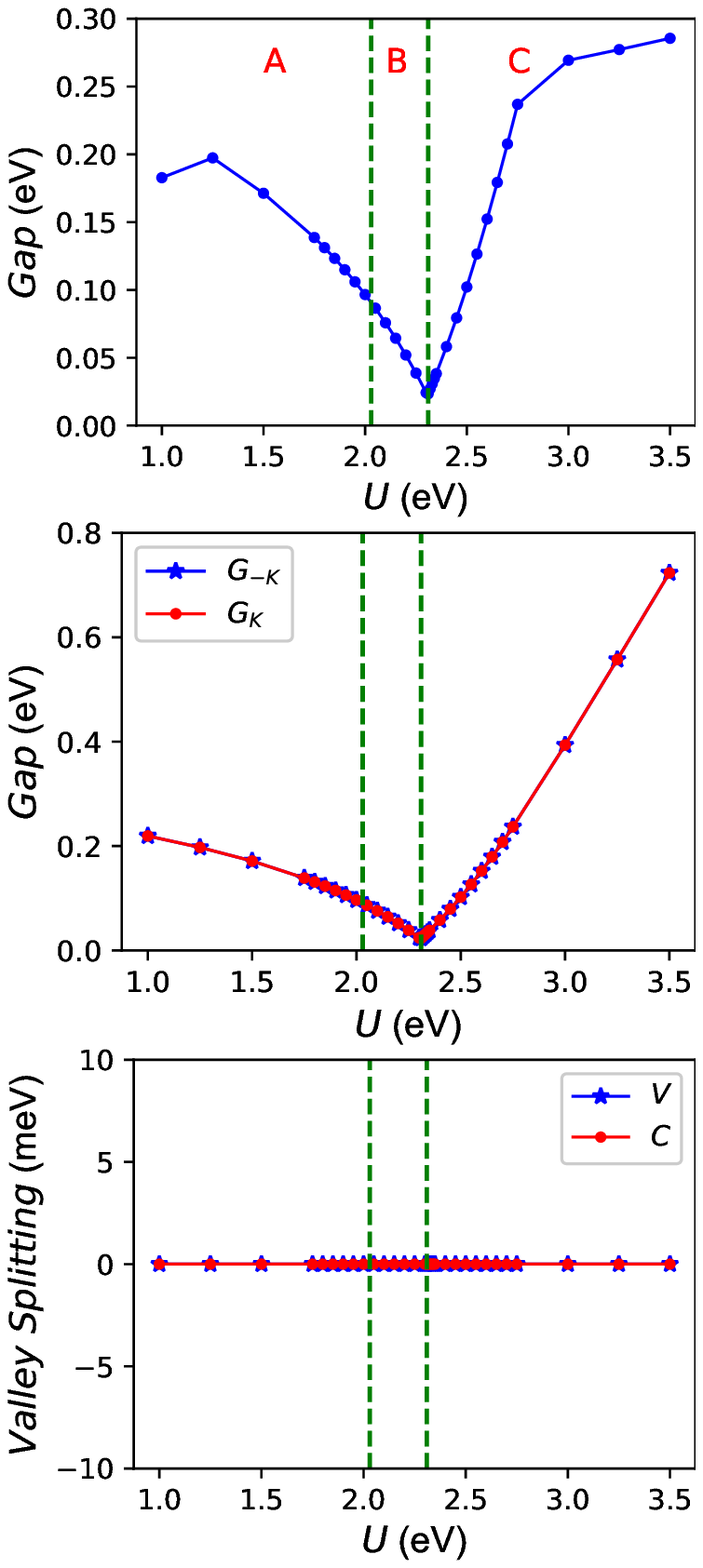}
  \caption{(Color online)For $\mathrm{VSi_2P_4}$ monolayer with fixed in-plane  MA, Top panel:  the  global energy band gap; Middle panel: the energy  band gaps for -K and K valleys; Bottom panel: the valley splitting for both valence and condition bands  as a function of   $U$. The A and C regions mean in-plane, while the B region for out-of-plane. }\label{gap-1}
\end{figure}

\begin{figure}
  \includegraphics[width=8cm]{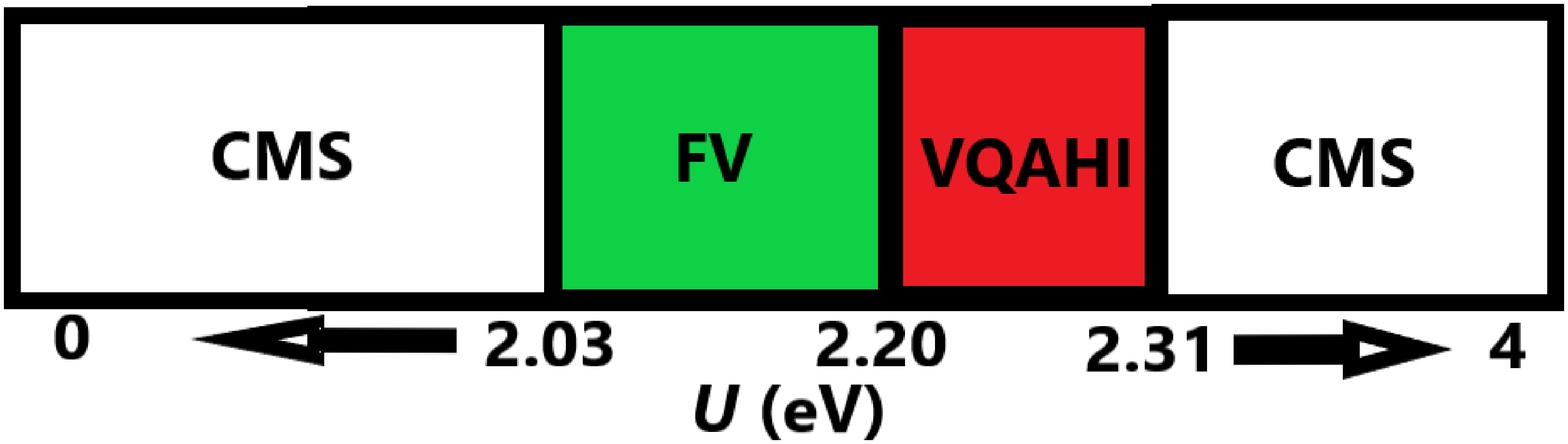}
  \caption{(Color online) The phase diagram for monolayer $\mathrm{VSi_2P_4}$ with different $U$ values, including CMS, FV and VQAHI. }\label{pha}
\end{figure}

\section{electronic structures}
 The magnetization is a pseudovector, which will lead to that the out-of-plane FM  breaks all
possible vertical mirror symmetry of the system, while  preserves the horizontal mirror symmetry.   The spontaneous valley polarization and a nonvanishing Chern number of  2D system can exist with the preserved horizontal mirror symmetry\cite{a4}. Thus, the MA of 2D system is very important to decide its electronic state.
 Firstly, the out-of-plane MA is fixed within considered $U$ range.
 The energy band structures of $\mathrm{VSi_2P_4}$ at some representative $U$ values  are plotted in FIG.3 of ESI. The evolutions of total energy band gap along with those at -K/K point and the valley splitting for both valence and condition bands  as a function of $U$ are shown  in \autoref{gap}.
Around  about $U$$=$2.20 eV and 2.40 eV, the total energy band gap of $\mathrm{VSi_2P_4}$  is closed, and the HVM  state can be achieved\cite{q10}, whose conduction electrons are intrinsically 100\% valley polarized.
At  about $U$$=$2.20 eV (2.40 eV),  the band gap of K (-K) valley  gets closed, while a  band gap at -K (K) valley is kept.

In a trigonal prismatic crystal field environment,  the  V-$d$ orbitals split
into low-lying $d_z^2$ orbital, $d_{xy}$+$d_{x^2-y^2}$  and
$d_{xz}$+$d_{yz}$ orbitals.
Calculated results show that $d_{x^2-y^2}$+$d_{xy}$ or $d_{z^2}$  orbitals of V atoms dominate K and -K valleys of  both  valence and conduction bands, and the V-$d$ orbital characters energy band structures at representative  $U$$=$1.50 eV, 2.25 eV  and 3.00 eV  are plotted in FIG.4 of ESI.
For $U$$<$2.20 eV,  the $d_{x^2-y^2}$+$d_{xy}$/$d_{z^2}$ orbitals dominate K and -K valleys of  valence/conduction  bands (For example $U$$=$1.50 eV).
  With increasing $U$  between 2.20 eV  and 2.40 eV, the band inversion  between $d_{xy}$+$d_{x^2-y^2}$ and $d_{z^2}$ orbitals at K valley can be observed (For example $U$$=$2.25 eV), which is accompanied by the first HVM state. When $U$ is larger than 2.40 eV, the band inversion  between $d_{xy}$+$d_{x^2-y^2}$ and $d_{z^2}$ orbitals  occurs at -K valley (For example $U$$=$3.00 eV), along with the second HVM state. These lead to that the distributions of  $d_{x^2-y^2}$+$d_{xy}$ and  $d_{z^2}$ orbitals of 0 eV$<$$U$$<$2.20 eV  are opposite to those of 2.40 eV$<$$U$$<$4 eV.

The distributions of Berry curvature are calculated, and they are  plotted in FIG.5 at representative  $U$$=$1.50 eV, 2.25 eV  and 3.00 eV.
For 0 eV$<$$U$$<$2.20 eV  and 2.40 eV$<$$U$$<$4 eV,     the opposite
signs and different magnitudes for Berry curvatures can be observed around -K and K valleys.
However,  for 2.20 eV$<$$U$$<$2.40 eV, the Berry curvatures  around -K and K valleys show the same signs and different magnitudes.
These indicate that the flipping  of the sign of Berry curvature will occur  at K or -K valley.
The positive Berry curvature  (For example $U$$=$1.50 eV)  changes into negative  one  (For example $U$$=$2.25 eV) at K valley, and then the negative  Berry curvature (For example $U$$=$2.25 eV) changes into positive one (For example $U$$=$3.00 eV) at -K valley.
The twice flipping  of the sign of Berry curvature is related with twice band inversion  between $d_{xy}$+$d_{x^2-y^2}$ and $d_{z^2}$ orbitals.

With increasing $U$, twice gap close, band inversion and flipping  of the sign of Berry curvature  suggest twice topological phase transition, and the QAH state may exist.
To confirm this, the edge states and  Chern number are calculated. Calculated results show that there are no nontrivial chiral edge states for 0 eV$<$$U$$<$2.20 eV  and 2.40 eV$<$$U$$<$4 eV. When   $U$ is between 2.20 eV  and 2.40 eV, the  $\mathrm{VSi_2P_4}$ is a QAH phase.
 The  edge states at representative  $U$$=$2.25 eV  are plotted in \autoref{s-s}, which shows a nontrivial chiral edge state connecting  the conduction bands and valence  bands, implying a QAH phase. The predicted  Chern number $C$=-1 by integrating the Berry curvature  within the first BZ.

For 0 eV$<$$U$$<$2.20 eV,  the valley splitting of valence band is observable, while the valley splitting of conduction band can be ignored.
 However, for 2.40 eV$<$$U$$<$4 eV, the opposite situation can be observed  with respect to the case of 0 eV$<$$U$$<$2.20 eV.
 For 2.20 eV$<$$U$$<$2.40 eV, the valley splitting of valence/condition bands decreases/increases from observable/ignored value to ignored/observable one with increasing $U$. In this region, $\mathrm{VSi_2P_4}$  is a VQAHI with   spontaneous valley splitting and
 chiral edge states. These can be understood by the distributions of  $d_{x^2-y^2}$+$d_{xy}$ and  $d_{z^2}$ orbitals. If $d_{x^2-y^2}$+$d_{xy}$/$d_{z^2}$ orbitals  dominate  -K and K valleys, the valley splitting will be observable/ignored.
  The SOC Hamiltonian only
involving  the interaction of the same spin states mainly contributes to valley polarization, which with out-of-plane magnetization can be  simplified as\cite{f6,v2,v3}:
\begin{equation}\label{m1}
\hat{H}^0_{SOC}=\alpha \hat{L}_z
\end{equation}
where   $\hat{L}_z$/$\alpha$ is to the $z$ component of  orbital angular momenta/coupling strength. For $d_{x^2-y^2}$+$d_{xy}$ and  $d_{z^2}$ orbitals and the wave vector symmetry at the K and -K valleys, the basis functions
are chosen as:
\begin{equation}\label{m2}
   \begin{array}{c}
|\phi^\tau>=\sqrt{\frac{1}{2}}(|d_{x^2-y^2}>+i\tau|d_{xy}>)\\
or\\
|\phi^\tau>=|d_{z^2}>
  \end{array}
\end{equation}
 The resulting energy level at K
or -K valley can  be defined as:
\begin{equation}\label{m3}
E^\tau=<\phi^\tau|\hat{H}^0_{SOC}|\phi^\tau>
\end{equation}
where the subscript  $\tau=\pm1$  refers to valley index.
If $d_{x^2-y^2}$+$d_{xy}$ orbitals  dominate  -K and K valleys, the valley splitting $|\Delta E|$  is given by:
\begin{equation}\label{m4}
|\Delta E|=|E^{K}-E^{-K}|=4\alpha
\end{equation}
If the -K and K valleys are mainly from $d_{z^2}$ orbitals, the valley splitting $|\Delta E|$
 is expressed  as:
\begin{equation}\label{m4}
|\Delta E|=|E^{K}-E^{-K}|=0
\end{equation}
Clearly, these results
agree with the difference of the valley polarizations in the
conduction and valence bands with different $U$ from DFT calculations.

\begin{figure}
  \includegraphics[width=8cm]{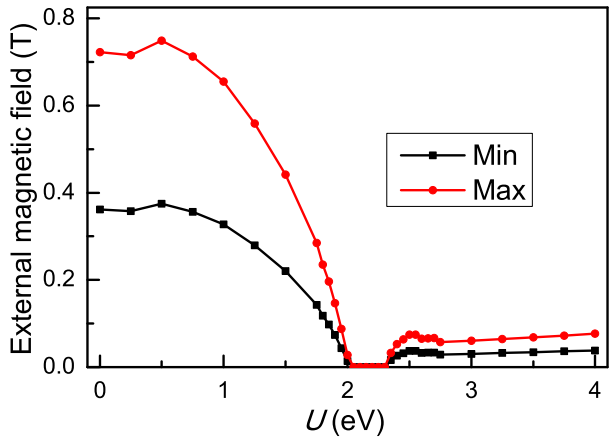}
  \caption{(Color online) The external magnetic field range  to overcome  energy barrier from in-plane to out-of-plane as a function of $U$. }\label{emf}
\end{figure}

Subsequently, the in-plane MA is assumed to investigate electronic properties of $\mathrm{VSi_2P_4}$ monolayer as a function of $U$.
The total energy band gap along with those at -K/K point and the valley splitting for both valence and condition bands  as a function of $U$ are
 plotted  in \autoref{gap-1}, and the representative energy band structures  are shown  in FIG.6 of ESI.
With increasing $U$, the gap  decreases, and  then increases. In considered $U$ range, the gaps at -K and K points coincide completely, and the valley  splitting is always zero. These mean that
no spontaneous valley polarization  and QAH phase can appear in $\mathrm{VSi_2P_4}$. For $d_{x^2-y^2}$+$d_{xy}$-dominated -K and K valley,   $\Delta E=4\alpha cos\theta$\cite{v3} with general magnetization orientation, where $\theta$=0/90$^{\circ}$ means out-of-plane/in-plane direction. For in-plane one, the valley splitting of $\mathrm{VSi_2P_4}$ become zero. Thus,  $\mathrm{VSiGeN_4}$ monolayer is a common magnetic semiconductor (CMS).

According to the MAE and electronic structure discussed above, the  intrinsic phase diagram of $\mathrm{VSi_2P_4}$ monolayer is shown in \autoref{pha}, and the electronic state includes CMS, FV and VQAHI. For $\mathrm{VA_2Z_4}$, the $U$=3 eV has been widely used in previous calculations\cite{a5,a12,a1-8,a1-9,a1-10,a1-10,a1-11,a1-12,a1-13}. The phase diagram shows that $\mathrm{VSi_2P_4}$ intrinsically is not a FV material.  However, the FV states can be easily achieved by external magnetic field due to small MAE.
The external magnetic field  to overcome  energy barrier from in-plane to out-of-plane as a function of $U$ is plotted in \autoref{emf}.
At $U$=3 eV, the   energy barrier  is equivalent to applying
a  external magnetic field of around 0.03-0.06 T. Although the $U$ can not be  directly tuned in experiment, it can be  regulated equivalently by strain, which has been confirmed in monolayer $\mathrm{RuBr_2}$\cite{a7}.  For a given material, the correlation strength should be fixed.
Even though $\mathrm{VSi_2P_4}$ dose not  belong to special QAH phase in the
phase diagram, the QAH state can be realized by strain or  external magnetic field and strain. Strain-induce QAH states have been predicted in many 2D systems\cite{a7,a1-11,a1-13,t8}.

\section{Conclusion}
In summary, we have demonstrated that the inclusion of MSA  can result in a different
phase diagram in $\mathrm{VSi_2P_4}$. For fixed out-of-plane situation, the  VQAHI state can be observed  between two HVM states, which is related with twice sign-reversible  Berry curvature and  twice band inversions of $d_{xy}$+$d_{x^2-y^2}$ and $d_{z^2}$ orbitals at -K or K valleys. For assumed in-plane situation, $\mathrm{VSi_2P_4}$ is a CMS, and  shows no spontaneous valley polarization.
At representative $U$$=$3 eV\cite{a5,a12,a1-8,a1-9,a1-10,a1-10,a1-11,a1-12,a1-13}, $\mathrm{VSi_2P_4}$ is a CMS,  which  is different from previous FV material with only considering MCA energy\cite{a5,a1-9}. However, spontaneous valley polarization  can be realized by small external magnetic field.
It is possible to achieve  QAH state in $\mathrm{VSi_2P_4}$  by strain or  external magnetic field and strain
 Our works can  deepen our understanding of MSA in the V-based 2D $\mathrm{MA_2Z_4}$ family materials.

\begin{acknowledgments}
This work is supported by the Natural Science Basis Research Plan in Shaanxi Province of China  (2021JM-456) and the National Natural Science Foundation of China (No. 12104130). We are grateful to Shanxi Supercomputing Center of China, and the calculations were performed on TianHe-2.
\end{acknowledgments}

\end{document}